\documentclass[aps,prl,reprint,superscriptaddress]{revtex4-2}

\usepackage[T1]{fontenc}
\usepackage[utf8]{inputenc}
\usepackage{bbold}
\usepackage{amsmath,amssymb,bm}
\usepackage{graphicx}
\usepackage{xcolor}

\definecolor{darkblue}{rgb}{0,0,0.6}
\definecolor{darkred}{rgb}{0.6,0,0}
\usepackage[colorlinks=true,urlcolor=darkblue, citecolor=darkblue, linkcolor=darkred, hyperfootnotes=false]{hyperref}

\newcommand{\dd}{\mathrm{d}}
\newcommand{\ee}{\mathrm{e}}
\newcommand{\avg}[1]{\left\langle #1 \right\rangle}

\newcommand{\abs}[1]{\left| #1 \right|}

\newcommand{\moy}[1]{\left\langle #1 \right\rangle}

\def\Tp{\mathrm{T}_+}

\newcommand{\I}{\mathrm i}

\newcommand{\1}{\mathbf 1}
\newcommand{\un}{\mathbb 1}

\newcommand{\sg}[1]{\mathrm{sign}({#1})}

\newcommand{\avgi}[1]{\left\langle #1 \right\rangle_{\mathrm{i}}}
\newcommand{\avge}[1]{\left\langle #1 \right\rangle_{\mathrm{e}}}

\def\Transf{\mathcal{R}}

\begin{document}

\title{Exact Microscopic One-Body Reduction in the Symmetric Exclusion Process}

\author{Olivier Bénichou}
\affiliation{Sorbonne Universit\'e, CNRS, Laboratoire de Physique Th\'eorique de la Mati\`ere Condens\'ee (LPTMC), 4 Place Jussieu, 75005 Paris, France}

\author{Aurélien Grabsch}
\affiliation{Sorbonne Universit\'e, CNRS, Laboratoire de Physique Th\'eorique de la Mati\`ere Condens\'ee (LPTMC), 4 Place Jussieu, 75005 Paris, France}

\begin{abstract}
The symmetric exclusion process (SEP) is a paradigmatic model of interacting particle transport. Its current--density correlations are governed by an infinite many-body hierarchy of equations. Only recently, this hierarchy was found to collapse at large scales onto a single closed equation. Is this remarkable simplification created by coarse graining, or does it reflect a deeper property already present in the microscopic dynamics? Here we show that the second scenario is realized. At arbitrary time on the lattice, the entire hierarchy collapses onto a single linear equation which is exactly equivalent to a persistence problem for a single auxiliary random walker. Thus an interacting many-body problem reduces exactly to a genuine one-body problem before any scaling limit is taken. This reduction determines the complete finite-time correlation profile, including during nonequilibrium relaxation.  More broadly, the microscopic operator uncovered here opens a route to more general correlation functions in the SEP, and may point to a general organizing principle for nonequilibrium interacting particle systems.
\end{abstract}

\maketitle

\paragraph{Introduction.---}
Single-file diffusion describes particles moving in narrow channels under the constraint that they cannot bypass one another. This simple geometrical constraint generates strong correlations and anomalous transport, making single file diffusion a fundamental setting for studying interacting particle dynamics~\cite{Harris:1965,Levitt:1973,Arratia:1983,Hahn:1996,Wei:2000,Kollmann:2003,Lin:2005}. Its simplest lattice realization is the symmetric exclusion process (SEP), a paradigmatic model of nonequilibrium statistical physics in which particles perform symmetric random walks subject to hard-core exclusion~\cite{Spitzer:1970,Golinelli:2006,Derrida:2007,Chou:2011,Mallick:2015}.

A central observable is the integrated current $Q_t$ through a fixed bond~\cite{Derrida:2025a}. Its statistics have been studied extensively~\cite{Derrida:2004,Appert:2008,Derrida:2009,Akkermans:2013,Saha:2023,Saha:2024,Grabsch:2024c,Berlioz:2024,Sharma:2026}, with its full distribution derived exactly by relying on Bethe ansatz for an initial density step~\cite{Derrida:2009}. The distribution of $Q_t$, however, does not reveal the spatial structure associated with a current fluctuation. This information is encoded in the correlation profile
\begin{equation}
 w_r(t;\lambda)=
 \frac{\avg{\eta_r(t)\ee^{\lambda Q_t}}}
 {\avg{\ee^{\lambda Q_t}}},
 \label{eq:wdef}
\end{equation}
whose expansion in $\lambda$ generates the correlations between the local density at position $r$, represented by the occupation number $\eta_r(t)=0$ or $1$, and the integrated current $Q_t$. Equivalently, $w_r$ is the mean density in the ensemble biased by the integrated current.

Determining these correlations is, a priori, a genuine many-body problem. The evolution of the one-point quantity $w_r$ involves two-point extensions of~\eqref{eq:wdef}, whose evolution in turn involves three-point functions, and so on. This generates an infinite hierarchy of equations, with no obvious mechanism by which it could close. Remarkably, only recently was it found that, at large scales, the hierarchy relevant to this profile closes into a single equation~\cite{Poncet:2021,Grabsch:2022,Grabsch:2023} of Wiener--Hopf type for the scaling behavior of $w_r$.

The simplicity of this result is especially striking because it does not follow from any obvious factorization of the microscopic dynamics.
The closure was first uncovered from the systematic study of successive orders in $\lambda$ of the correlation profile~\eqref{eq:wdef} at large scale~\cite{Grabsch:2022}. The resulting closed Wiener--Hopf equation was later proved by Mallick, Moriya, and Sasamoto~\cite{Mallick:2022}, who applied the inverse scattering method of classical integrability~\cite{Ablowitz:1981} to the equations of macroscopic fluctuation theory (MFT) describing the SEP at large scale~\cite{Bertini:2015}.
Related Wiener--Hopf structures also occur in other transport models~\cite{Bettelheim:2022,Bettelheim:2022a} and in the KPZ equation~\cite{Krajenbrink:2021,Krajenbrink:2022a,Krajenbrink:2023a}.

This raises a fundamental question about the origin of this unexpected simplification. There are two sharply different possibilities. The closure may be an emergent property at large scale: coarse graining and diffusive scaling could erase the complexity of the microscopic hierarchy and produce a much simpler effective equation, with no exact lattice counterpart at finite time. Alternatively, the closed equation may be the large-scale manifestation of a stronger identity already present in the microscopic dynamics, before any scaling limit is taken.

Distinguishing between these two scenarios is nontrivial. Exact finite-time microscopic formulas are known for the  current statistics~\cite{Derrida:2009}, but they do not provide an equation for the spatially resolved profile~\eqref{eq:wdef}. Conversely, the inverse scattering construction establishes the closure only after passing to the continuum description~\cite{Mallick:2022}.
Remarkably, the very possibility that a closure might already exist at the microscopic level, rather than emerge from coarse graining, has, to our knowledge, never even been raised.

Here we show that the second scenario is realized. At arbitrary time and
directly on the lattice, the entire hierarchy governing the correlation
profile collapses onto a single scalar linear equation. We further show that
this equation can be interpreted in terms of a single auxiliary random walker.
Thus the spatial structure associated with current fluctuations in an interacting many-body system reduces exactly to a genuine one-body problem before any scaling limit is taken.

Our derivation combines the stirring representation of the SEP~\cite{Arratia:1983}
with an extension of the microscopic Bethe-ansatz calculation used for the
current statistics in Ref.~\cite{Derrida:2009}.
Our result thus reveals a striking and unexpected feature of many-body dynamics: the spatial organization of current fluctuations in a paradigmatic interacting system is governed, down to the microscopic scale and even far from equilibrium, by a hidden one-body structure.

\paragraph{Main result: exact microscopic closure.---}
We consider the SEP on the infinite lattice $\mathbb{Z}$. Each particle attempts to jump to either neighboring site with unit rate, with jumps to occupied sites rejected by the exclusion rule. We denote by $\eta_r(t)\in\{0,1\}$ the occupation of site $r$ at time $t$. Initially, sites $r\leq0$ are independently occupied with probability $\rho_-\in[0,1]$, while sites $r>0$ are occupied with probability $\rho_+$.

We write $\nabla f_r=f_{r+1}-f_r$ and introduce the fixed linear transform
\begin{equation}
    (\Transf f)_s
    =
    \sum_{m=1}^{s}(-1)^{m-1}
    \binom{s-1}{m-1}2^{s-m}f_m
    \:.
\label{eq:DefTransf}
\end{equation}
We then define the normalized gradients $\mathcal{W}_r$ of the physical profile~\eqref{eq:wdef} by
\begin{equation}
\label{eq:WdefMain}
\mathcal{W}_r(t;\lambda)=
\begin{cases}
-\displaystyle\frac{\nabla w_r(t;\lambda)}
{\frac{1}{1-\ee^{-\lambda}}-w_1(t;\lambda)},
& r\geq1,\\[4mm]
\displaystyle\frac{\nabla w_r(t;\lambda)}
{\frac{1}{1-\ee^{\lambda}}-w_0(t;\lambda)},
& r\leq-1.
\end{cases}
\end{equation}
For $r\geq1$, we set $\mathcal W_r^{(+)}=\mathcal W_r$ and $\mathcal W_r^{(-)}=\mathcal W_{-r}$. Our main result is that the two branches obey the same closed equation,
\begin{equation}
\label{eq:ClosedMicro}
\boxed{
\mathcal W_r^{(\pm)}(t;\lambda)
+\omega\sum_{s\geq1}p_{r+s}(t)
\bigl(\Transf \mathcal{W}^{(\pm)}\bigr)_s
=\omega p_r(t)
\:,
}
\end{equation}
where $p_n(t) = \ee^{-2t}\mathrm I_n(2t)$ is the free one-particle propagator, with $\mathrm I_n$ the modified Bessel function, while
\begin{equation}
\label{eq:defomega}
\omega=
[1+\rho_-(\ee^\lambda-1)]
[1+\rho_+(\ee^{-\lambda}-1)]
-1
\end{equation}
is the usual parameter controlling the SEP current statistics~\cite{Derrida:2004,Derrida:2009}.

Equation~\eqref{eq:ClosedMicro} displays the microscopic closure in its strongest form. It is a single linear equation for each half-line, and contains no higher-order correlation function: the entire many-body hierarchy has disappeared. Moreover, it involves only the free one-particle propagator $p_n(t)$. The dependence of the closed equation on the initial densities and on the parameter $\lambda$ is reduced to the single scalar parameter $\omega$. Thus the spatial correlation problem of the interacting SEP is reduced exactly, at arbitrary time, to a linear problem.

The normalization~\eqref{eq:WdefMain} brings the microscopic closure to the compact form~\eqref{eq:ClosedMicro}. Remarkably, the structure of the hydrodynamic Wiener--Hopf equation of Ref.~\cite{Grabsch:2022} is already present at the microscopic level, with the known equation recovered directly in the large-scale  limit, as we now show.

\paragraph{Recovering the large-scale closure.---}
At large scales, the correlation profile admits the diffusive scaling~\cite{Poncet:2021}
\begin{equation}
    \label{eq:ScalingFormW}
    w_{r}(t;\lambda)
    \underset{t \to \infty}{\simeq}
    \Phi\left(
    x \equiv \frac{r}{\sqrt{t}};\lambda
    \right)
    \:.
\end{equation}
Under the same scaling, the free lattice propagator $p_n(t)$ converges to the Gaussian kernel,
\begin{equation}
p_{\lfloor x\sqrt t\rfloor}(t)
\simeq
\frac{1}{\sqrt t}\,G(x),
\qquad
G(x)=\frac{\ee^{-x^2/4}}{\sqrt{4\pi}}.
\end{equation}
From these scalings, for  $x>0$,
\begin{equation}
\label{eq:WtoOmegaMain}
\sqrt{t} \mathcal W_{\lfloor x\sqrt{t}\rfloor}^{(\pm)}(t;\lambda)
\simeq
\Omega^{(\pm)}(\pm x;\lambda)
\equiv
\frac{\mp \Phi'(\pm x;\lambda)}{\frac{1}{1-\ee^{ \mp \lambda}} - \Phi(0^\pm;\lambda)}
\:.
\end{equation}

The distinctive microscopic ingredient is the transform $\Transf$. To determine its large scale behavior, we introduce the
translation operator $\Tp f_r=f_{r+1}$.
The action of $\Transf$~\eqref{eq:DefTransf} can thus be written as
$(\Transf f)_s=(2-\Tp)^{s-1}f_1$.
At large scale, the translation operator becomes $T_+ \simeq \exp(t^{-1/2}\partial_x)$, hence $(2-\Tp)^{\lfloor z \sqrt{t} \rfloor - 1} \simeq \ee^{-z \partial_x}$, so that
\begin{equation}
    \label{eq:ReflecMacro}
   \sqrt{t} \big( \Transf \mathcal{W}^{(+)} \big)_{\lfloor z \sqrt{t} \rfloor}
   \underset{t \to \infty}{\simeq}
   \Omega^{(+)}(-z;\lambda)
   \:.
\end{equation}
Thus the lattice transform $\Transf$ becomes a reflection across the
origin: it evaluates the continuation of a half-line profile at the
reflected coordinate.
Inserting these scaling forms into the closed equation~\eqref{eq:ClosedMicro}, we obtain for $x \gtrless 0$,
\begin{equation}
\label{eq:ClosedMacro}
\Omega^{(\pm)}(x)
+\omega\int_0^\infty
\Omega^{(\pm)}(\mp z)\,G(x\pm z)\,\dd z
=\omega G(x)
\:.
\end{equation}
These are precisely the Wiener--Hopf equations
found in Ref.~\cite{Grabsch:2022} and later proved by Mallick, Moriya,
and Sasamoto~\cite{Mallick:2022} from the integrability of the MFT
equations.
The large-scale closure therefore follows directly from the
microscopic equation~\eqref{eq:ClosedMicro} by diffusive scaling.

\paragraph{Discussion.---}
Equation~\eqref{eq:ClosedMicro} has several immediate consequences.
First, the exact microscopic equation~\eqref{eq:ClosedMicro} remarkably takes the same form as the large-scale equation~\eqref{eq:ClosedMacro}.
The discrete transform $\Transf$ implements at the level of the lattice the reflection~\eqref{eq:ReflecMacro}, without the need of any continuation to $r \leq 0$.
This becomes clear upon considering the Poisson transform
\begin{equation}
\label{eq:PoissonTransf}
(\mathcal{P} f)(x)
\equiv
\ee^{-x}\sum_{r\ge1}f_r\frac{x^{r-1}}{(r-1)!}\:,
\end{equation}
which after summation of~\eqref{eq:DefTransf} yields
\begin{equation}
\label{eq:Creflection}
\mathcal{P}(\Transf f)(x)=\mathcal{P} f(-x)
\:.
\end{equation}
Thus $\Transf$ is exactly the reflection $x\mapsto -x$ in the
Poisson representation. The microscopic interaction is therefore
encoded by a simple reflection, echoing the hard-core constraint at contact
in the SEP.

Second, the equation~\eqref{eq:ClosedMicro} being linear, it can be solved analytically, as discussed below, or numerically.

Third, Eq.~\eqref{eq:ClosedMicro} shows that the many body problem of the current in the SEP can be reduced to a single body problem, as we now proceed to show.

\paragraph{A one body reduction.---}
We now show that the closed microscopic equation~\eqref{eq:ClosedMicro} describing the many-body dynamics of the SEP actually describes a single degree of freedom.
For this, we consider the Poisson transform~\eqref{eq:PoissonTransf} of the closed equation~\eqref{eq:ClosedMicro}. To evaluate the transform of the sum, we use the definition of the Bessel function $\mathrm{I}_n(2t) = \sum_{k\geq 0} \frac{t^{2k+n}}{k! (n+k)!}$~\cite{Gradshteyn:1994} to derive
\begin{align}
    \mathcal{P}[r \mapsto p_{s+r}(t)](T)
    &=
    \ee^{-T-2t}\partial_T^s
    \sum_{k=0}^\infty \frac{t^{k+1} (t+T)^k}{k! (k+1)!}
    \\
    &= \int_0^t
    G(T+t|t-u)
     \frac{u^{s-1} \ee^{-u}}{(s-1)!}
    \dd u
    \nonumber
    \:,
\end{align}
where we introduced
\begin{align}
    \label{eq:TrProb}
    G(\tau|\tau')
    &=
    \ee^{-\tau-\tau'}\sum_{k=0}^\infty  \frac{\tau^k (\tau')^{k+1}}{k! (k+1)!}
    \\
    &=
    \ee^{-\tau-\tau'}
    \sqrt{\frac{\tau'}{\tau}} \:
    \mathrm{I}_1  \left(
    2 \sqrt{\tau \tau'}
    \right)
    \nonumber
    \:.
\end{align}
We thus obtain the Poisson transform of~\eqref{eq:ClosedMicro} as
\begin{equation}
    \label{eq:PoissonClosed}
    \psi_t(T)
    + \omega
    \int_0^t
    G(T|u)
    \psi_t(u)
    \dd u
    = \omega \: G(T|t)
    \:,
\end{equation}
where we defined $\psi_t(T) = (\mathcal{P}\mathcal{W}^{(+)})(T-t)$ and used the reflection property~\eqref{eq:Creflection}.
Equation~\eqref{eq:PoissonClosed} is equivalent to the closed equation~\eqref{eq:ClosedMicro} and directly admits a single-body interpretation. Indeed, define a random walk in discrete time $X_n$, $n \in \mathbb{N}$, with $X_0 = t$ and transition probability density from $X_n$
 to $X_{n+1}$ given by  $G(X_{n}|X_{n+1})$. Let $q_{n,t}(T)$ denote the probability density that the walker has remained below $t$ for the first $n-1$ steps and is located at $T$ at step $n$. Using the identity $G(T|u)=(u/T)G(u|T)$,
Eq.~\eqref{eq:PoissonClosed} is equivalent to
the renewal equation satisfied by the generating
function of the probability densities $q_{n,t}$,
 \begin{equation}
    -\frac{T}{t} \psi_t(T)
     = \sum_{n=1}^\infty (-\omega)^n q_{n,t}(T)
     \:.
 \end{equation}
Thus, the closed microscopic equation~\eqref{eq:ClosedMicro} reduces the many-body problem of current--density correlations in the SEP to the study of a single random walker.

 Note that, in the large scale limit, the transition probability reduces to
 \begin{equation}
     G(t + 2 \sqrt{t}x| t + 2 \sqrt{t}y)
     \underset{t \to \infty}{\simeq}
     \frac{\ee^{-(x-y)^2}}{2 \sqrt{\pi t}}
     \:,
 \end{equation}
 recovering the Gaussian kernel entering the large-scale closure of~\cite{Grabsch:2022}. It is remarkable that the same one body reduction holds at the microscopic level, with transition probability~\eqref{eq:TrProb}.

\paragraph{Exact finite-time solution and asymptotics.---} Let us write the closed microscopic equations~\eqref{eq:ClosedMicro} as $[M\mathcal{W}^{(\pm)}]_r = \omega p_r$, with the matrix of entries
\begin{equation}
    M_{r,m} = \delta_{r,m}+\omega
    (-1)^{m-1} \sum_{s=0}^{\infty}
    \binom{s+m-1}{m-1} 2^{s}
    p_{r+s+m}
    \:,
\end{equation}
for $r,m \geq 1$.
The system of linear equations can thus be solved using Cramer's rule, which gives
\begin{equation}
    \mathcal{W}_r^{(+)}
    =  \mathcal{W}_r^{(-)}
    = \omega \frac{\det(\widetilde{M}_r)}{\det(M)}
    \:,
\end{equation}
in terms of the infinite determinants of $M$ and $\widetilde{M}_r$ which corresponds to $M$ with the column $r$ replaced by $\{ p_s \}_{s \geq 1}$. To get the expression of $w_r$, we use the multilinearity of the determinant to write $\det[M_{i,j} + \omega p_i \un_{j \leq r}]_{i,j \geq 1} = \det(M) + \omega \sum_{j=1}^{r} \det(\widetilde{M}_j)$, since the determinant vanishes when two columns are equal. Using this relation with the definition of $\mathcal{W}_r^{(+)}$~\eqref{eq:WdefMain} together with $\lim_{r \to \infty} w_r = \rho_+$, we obtain a representation of the correlation profile~\eqref{eq:wdef} for $r \geq 1$ as a ratio of determinants
\begin{equation}
    w_r = C_+
    - \left(
    C_+
    - \rho_+
    \right)
    \frac{\det[M_{ij} + \omega p_i \un_{j < r}]_{i,j \geq 1}}{\det[M_{ij} + \omega p_i]_{i,j \geq 1}}
    \:,
\end{equation}
with $C_+ = (1-\ee^{-\lambda})^{-1}$.
This provides an exact solution for the correlation profile at all time.

The microscopic closure also yields fully explicit
short-time and long-time expansions of the correlation
profile at fixed lattice sites. Here we focus on a
particularly revealing consequence: the first finite-time
correction to the long time scaling limit.

At fixed $x=|r-\frac12|/\sqrt{t}>0$, let
$\Phi_\pm(x)$ denote the known large-scale profile, with the upper (lower) sign corresponding to $r\geq1$ ($r\leq0$). We obtain
\begin{align}
w_r(t;\lambda)
&=\Phi_\pm(x)+\frac{\Phi_{\pm,1}(x)}{\sqrt{t}}
+O(t^{-1}) \:,
\\
\Phi_{\pm,1}(x)
&=-\frac{a_1}{2}[C_\pm-\Phi_\pm(x)]
- \frac{1}{4}\int_0^x y\Phi_\pm'(y)\,\dd y
\nonumber\\
&\quad-\frac14\bigg[
(2+x^2)\Phi_\pm'(x)
+2x\Phi_\pm''(x)
\bigg]
\:,
\label{eq:HydroCorrection}
\end{align}
where $C_\pm=(1-\ee^{\mp\lambda})^{-1}$ and
$a_1=-\operatorname{Li}_{1/2}(-\omega)/(2\sqrt{\pi})$.

Remarkably, the full correction is expressed explicitly in terms of the known large-scale profile, without solving any additional Wiener--Hopf equation.

\begin{figure}
    \centering
    \includegraphics[width=0.7\columnwidth]{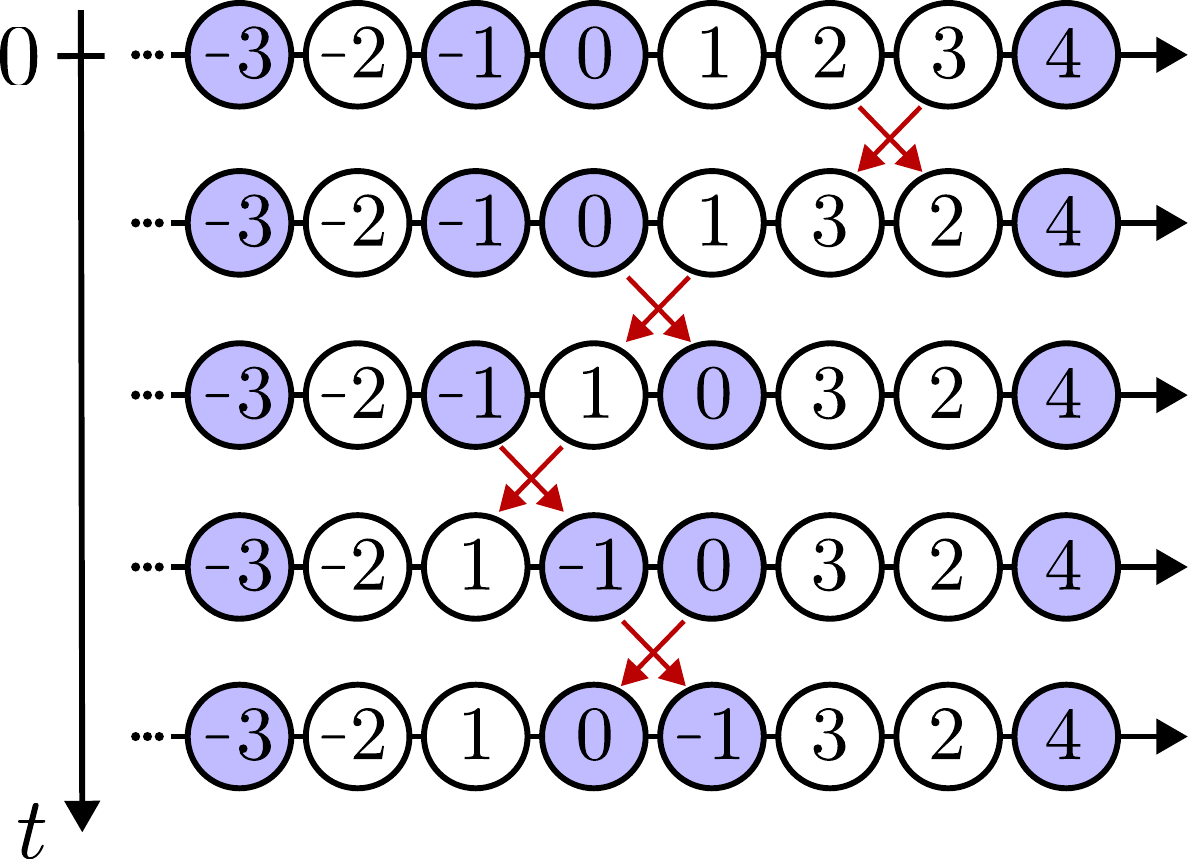}
    \caption{Representation of the SEP as a stirring process~\cite{Arratia:1983}. At random times, picked with unit rate, two neighboring integers are exchanged (red arrows). A configuration of the SEP is obtained by filling the sites hosting a label that was initially occupied by a particle (blue). Note that exchanging the labels of two occupied sites yield the same SEP configuration.}
    \label{fig:SEPStirring}
\end{figure}

\paragraph{Derivation of the microscopic closed equations.---}
The first step to derive the closed equations for $w_r$~\eqref{eq:ClosedMicro} is to reduce to a universal correlation profile, for any step of density $(\rho_-,\rho_+)$, by following the approach of Arratia~\cite{Arratia:1983}.
We decompose the averaging over the realizations of the SEP in~\eqref{eq:wdef} into an average $\avgi{\cdots}$ over the initial condition and an average $\avge{\cdots}$ over the evolution of a stirring process on all the integers in $\mathbb{Z}$, which is generated by exchanging neighboring integers with unit rate. The configuration of the SEP at any time is deduced from the one of the stirring process by following the labels of the sites that are initially occupied, as shown in Fig.~\ref{fig:SEPStirring}.
More precisely, denoting $n(i,t) \in \mathbb{Z}$ the position of integer $i \in \mathbb{Z}$ at time $t$ in the stirring process, the site $n(i,t)$ in the SEP is occupied if and only if $\eta_i(0) = 1$. This allows us to express the integrated current as the contribution of the negative occupied labels that moved to the positive axis, minus the contribution of positive ones,
\begin{equation}
    Q_t = \sum_{i \leq 0} \eta_i(0) \un_{n(i,t) > 0}
    - \sum_{i > 0} \eta_i(0) \un_{n(i,t) \leq 0}
    \:.
\end{equation}
Since both $\eta_i$ and $\un$ take only values $0$ or $1$, we can write
\begin{multline}
    \ee^{\lambda Q_t} = \prod_{i \leq 0} \left[
    1 + (\ee^{\lambda} - 1) \eta_i(0)\un_{n(i,t) > 0}
    \right]
    \\
    \times \prod_{i > 0} \left[
    1 + (\ee^{-\lambda} - 1) \eta_i(0)\un_{n(i,t) \leq 0}
    \right]
    \:.
\end{multline}
The occupations at initial time being independent, we can easily average over the initial realizations to get
\begin{equation}
    \avg{\ee^{\lambda Q_t}}
    = \avge{ \avgi{ \ee^{\lambda Q_t} } }
    = \avge{(1+\omega)^{N_t}}
    \:,
\end{equation}
where $N_t = \sum_{i>0} \un_{n(i,t) \leq 0} = \sum_{i \leq 0} \un_{n(i,t) > 0}$ is the number of negative integers that have reached the positive lattice at time $t$ in the stirring process, and $\omega$ is given by~\eqref{eq:defomega}. This directly proves the $\omega$ dependence of the moment generating function of $Q_t$ in a transparent way, compared to the original derivation~\cite{Derrida:2009}.
Proceeding similarly with the correlation profile~\eqref{eq:wdef}, we get
\begin{equation}
\label{eq:defhr}
    w_r(t; \lambda) = \left\lbrace
    \begin{array}{ll}
        \rho_+
        - \frac{\omega}{(\ee^{-\lambda}-1)A_-} h_r(t;\omega)
        & \text{for } r > 0 \:, \\
        \rho_- - \frac{\omega}{(\ee^{\lambda}-1)A_+} h_{1-r}(t;\omega) & \text{for } r \leq 0 \:,
    \end{array}
    \right.
\end{equation}
with $A_\pm = 1 + \rho_\pm(\ee^{\mp \lambda}-1)$ and
\begin{equation}
    h_r(t;\omega) = \sum_{j \leq 0} \frac{\avge{ \delta_{r,n(j,t)} (1+\omega)^{N_t}}}{\avge{(1+\omega)^{N_t}}}
\end{equation}
is the equivalent of the correlation~\eqref{eq:wdef} for the stirring process. The same function $h_r$ determines $w_r$ for any choice of the initial densities $(\rho_-,\rho_+)$. In the following, we thus focus on determining $h_r$. Note that $h_r$ has a simple meaning since for $\rho_- = 1$ and $\rho_+ = 0$, we have $w_r = h_r = 1-h_{1-r}$. We can thus focus on determining the correlation profile for $(\rho_- = 1,\rho_+ = 0)$ and deduce the profile for any density using~\eqref{eq:defhr}.

We can now turn to the second step of the derivation, which consists in determining the universal profile $h_r$.
For the specific choice $(\rho_- = 1,\rho_+ = 0)$, we have $\omega = \ee^{\lambda} - 1$ and $Q_t = \sum_{n > 0} \eta_n(t)$. Therefore, for $r \leq 0$, $ \avg{ \eta_r(t) \ee^{\lambda Q_t} }
    = \sum_{n \geq 1} \omega^{n-1} T_{n,r}$ where
\begin{equation}
    \label{eq:defTnr}
    T_{n,r} =
    \sum_{0 < r_2 < \cdots < r_n}
    \moy{ \eta_r(t) \eta_{r_2}(t) \cdots \eta_{r_n}(t) }
    \:.
\end{equation}
The expression for $r>0$ is more difficult to obtain, since it would involve a sum over ordered indices different from $r$. Nevertheless, it can be deduced from the particle-hole symmetry of the SEP, so we can restrict ourselves to the simpler $r \leq 0$ case. The multi-point correlation can be computed using the $n$-particle propagator of the SEP, obtained from Bethe ansatz in~\cite{Derrida:2009}. We show in the Appendix A how the derivation used in~\cite{Derrida:2009} can be adapted to account for the additional term $\eta_r$ in~\eqref{eq:defTnr}. We finally obtain that $h_r$ is expressed as a Fredholm determinant
\begin{equation}
    \label{eq:hrAsDet}
    h_r(t;\omega) = - \frac{\dd}{\dd s}
    \ln \det[ 1 + \omega \mathrm{K}_t + s b_{r;t} a^T] \Big|_{s=0}
    \:,
\end{equation}
where $\mathrm{K}_t$ is the usual kernel of the SEP~\cite{Derrida:2009}
\begin{equation}
    \label{eq:KernelSEP}
    \mathrm{K}_t(z,z') = \frac{\ee^{t \left( z + \frac{1}{z} -2 \right)}}{z z' + 1 -2z}
    \:,
\end{equation}
for $z$ and $z'$ taken on a small contour around the origin, and we defined the vectors
\begin{equation}
    \label{eq:VectorsAB}
    a(z) = \frac{1}{z-1}
    \:,
    \quad
    b_{r;t}(z) = z^{r-1} \ee^{t \left( z + \frac{1}{z} -2 \right)}
    \:.
\end{equation}
From Jacobi's formula on the derivative of the determinant, we can equivalently write
$
    h_r(t;\omega) = - \mathrm{Tr}\left[
    (1+\omega \mathrm{K}_t)^{-1} b_{r;t} a^T
    \right]
    = \phi_{t;\omega}^T b_{r;t}
$
where $\phi_{t;\omega} \equiv -(1+\omega \mathrm{K}_t^{T})^{-1} a$. Explicitly,
\begin{equation}
    \label{eq:Exprhr}
    h_r(t;\omega) = \oint \frac{\dd z}{2 \I \pi}
    z^{r-1} \ee^{t \left( z + \frac{1}{z} -2 \right)}
    \phi_{t;\omega}(z)
    \:,
    \quad
    r \leq 0
    \:,
\end{equation}
where $\phi$ is solution of the integral equation
\begin{equation}
    \label{eq:Exprphi}
    \phi_{t;\omega}(z)
    + \omega \oint \frac{\dd z'}{2\I \pi}
    \mathrm{K}_t(z',z) \phi_{t;\omega}(z')
    = \frac{1}{1-z}
    \:.
\end{equation}
These equations fully determine the function $h_r = 1 - h_{1-r}$ for $r \leq 0$, and thus the correlation profile $w_r$ for all $r$ via~\eqref{eq:defhr}.

The last step is to show from the explicit expressions~\eqref{eq:Exprhr}-\eqref{eq:Exprphi} that  the normalized profile $\mathcal W_r$ defined in~\eqref{eq:WdefMain} obeys the closed equations~\eqref{eq:ClosedMicro}.
We multiply the equation for $\phi$~\eqref{eq:Exprphi} by $(1-z)z^{r-1} \ee^{t \left(z + \frac{1}{z}-2 \right)}$ and integrate over $z$ on a small contour around the origin. This gives, for $r < 0$,
\begin{multline}
\label{eq:DblInteg0}
    \omega \oint \frac{\dd z \dd z'}{(2 \I \pi)^2} \frac{\ee^{t \left(z + \frac{1}{z}-2 \right)}
    \ee^{t \left(z' + \frac{1}{z'}-2 \right)}}{z z' + 1 - 2z'}
    z^{r-1}(1-z)
    \phi_{t;\omega}(z')
    \\
    = \nabla h_{r} + p_r
    \:.
\end{multline}
Computing now the double integral using~\eqref{eq:IdentityProlong} derived in Appendix B and performing a summation by parts, we obtain for $r < 0$,
\begin{equation}
    - \omega p_{r} h_0
    - \omega \sum_{s \geq 1} p_{s-r}
    (\Transf \nabla h^{(-)})_s
    =
    \nabla h_{r} + p_r
    \:,
\end{equation}
with $h^{(-)}_m = h_{-m}$.
This provides a closed equation for $h_r$, from which~\eqref{eq:defhr} and~\eqref{eq:WdefMain} give the normalized closure~\eqref{eq:ClosedMicro}.

\paragraph{Conclusion.---}
The central result of this work is the identification of the microscopic structure underlying the remarkable closure of current–density correlations in the SEP. At arbitrary time, including during nonequilibrium relaxation, the infinite correlation hierarchy reduces to free one-particle propagation coupled to a fixed operator that becomes a simple reflection after a Poisson transform. The resulting equation also admits an exact one-particle persistence interpretation, and its diffusive limit recovers the previously known large-scale Wiener–Hopf equation.

Thus, the large-scale closure is not a simplification generated by coarse graining: it is the continuum manifestation of an exact property of the microscopic dynamics. Beyond providing finite-time correlation profiles, our result identifies the operator responsible for this reduction. This offers a concrete framework for investigating further spatially resolved correlations in the SEP, and a precise criterion for testing whether analogous reductions survive for more complex observables.

\paragraph{Acknowledgments.---} We acknowledge support from the European Research Council (ERC) Synergy Grant No. 101224739 (UniCIPS). We thank P. Illien and P. Rizkallah for useful discussions on the stirring process.

% \bibliographystyle{apsrev4-1}
% \bibliography{bib}

%merlin.mbs apsrev4-1.bst 2010-07-25 4.21a (PWD, AO, DPC) hacked
%Control: key (0)
%Control: author (72) initials jnrlst
%Control: editor formatted (1) identically to author
%Control: production of article title (-1) disabled
%Control: page (0) single
%Control: year (1) truncated
%Control: production of eprint (0) enabled
%

\appendix
\setcounter{equation}{0}
\setcounter{figure}{0}
\setcounter{table}{0}
\renewcommand{\theequation}{A\arabic{equation}}

\onecolumngrid

\vspace{12pt}
\noindent\hrulefill \hspace{24pt} {\bf End Matter} \hspace{24pt} \hrulefill
\vspace{12pt}

\textit{Appendix A: Exact expression of the correlation profile.---} In this Section, we show how the sums~\eqref{eq:defTnr} can be computed to give the determinantal representation~\eqref{eq:hrAsDet} and thus the integral expression~\eqref{eq:Exprhr} from which the closed equation for $w_r$~\eqref{eq:ClosedMicro} is obtained.
Following~\cite{Derrida:2009}, we express the average of the product of occupation numbers in terms of the $n$-particle propagator $P_t^{(n)}(\mathbf{y}|\mathbf{x})$,
\begin{equation}
\label{eq:Tnrsums}
    T_{n,r}
    =
    \sum_{0< r_2 < \cdots < r_n}
    \moy{ \eta_r(t) \eta_{r_2}(t) \cdots \eta_{r_n}(t) }
    =
    \sum_{0< r_2 < \cdots < r_n}
    \sum_{x_1 < \cdots < x_n \leq 0}
    P_t^{(n)}(r, r_2, \ldots, r_n | x_1, \ldots, x_n)
    \:,
\end{equation}
where $P_t^{(n)}$ has an explicit Bethe ansatz expression
\begin{equation}
\label{eq:PropagBethe}
    P_t^{(n)}(\mathbf{y}|\mathbf{x})
    = \sum_\sigma \sg{\sigma}
    \oint \left[
    \prod_{k=1}^n \frac{\dd z_k}{2\I \pi z_k} \ee^{t \left( z_k + \frac{1}{z_k} - 2 \right)}
    z_{\sigma(k)}^{y_k - x_{\sigma(k)}}
    \right]
    \left[
    \prod_{k < l}
    \frac{z_{\sigma(k)} z_{\sigma(l)} + 1 - 2z_{\sigma(k)}}{z_k z_l + 1 - 2 z_k}
    \right]
    \:,
\end{equation}
where the integrals runs on small counterclockwise contours around the origin.
The ordered sums can be performed since they are geometric series
\begin{multline}
    T_{n,r} =
    \sum_\sigma \sg{\sigma}
    \oint \left[
    \prod_{k=1}^n \frac{\dd z_k}{2 \I \pi z_k} \ee^{t \left( z_k + \frac{1}{z_k} - 2 \right)}
    \right]
    z_{\sigma(1)}^r
    \left[ \prod_{k=2}^n
    \frac{z_{\sigma(k)} \cdots z_{\sigma(n)}}{1 - z_{\sigma(k) \cdots z_{\sigma(n)}}}
    \right]
    \\
    \times
    \left[
    \frac{1}{z_1 \cdots z_n}
    \prod_{k=1}^n
    \frac{z_1 \cdots z_k}{1 - z_1 \cdots z_k}
    \right]
    \left[
    \prod_{k < l}
    \frac{z_{\sigma(k)} z_{\sigma(l)} + 1 - 2z_{\sigma(k)}}{z_k z_l + 1 - 2 z_k}
    \right]
    \:.
\end{multline}
The goal is now to simplify these expressions using the formula derived by Tracy and Widom in~\cite{Tracy:2008} (and applied in~\cite{Derrida:2009}),
\begin{equation}
    \label{eq:TracyWidom1}
    \sum_\sigma \sg{\sigma}
    \left[
    \prod_{k=1}^n \frac{z_{\sigma(k)} \cdots z_{\sigma(n)}}{1 - z_{\sigma(k)} \cdots z_{\sigma(n)}}
    \right]
    \left[
    \prod_{k < l } \left(
    z_{\sigma(k)} z_{\sigma(l)}
    + 1 - 2 z_{\sigma(k)}
    \right)
    \right]
    =
    z_1 \cdots z_n \frac{\prod_{k< l} (z_l - z_k)}{ \prod_k (1-z_k)}
    \:.
\end{equation}
Unlike in~\cite{Derrida:2009}, we cannot apply this relation directly due to the site $r$ that is not summed over. We thus denote $j = \sigma(1)$ and apply the formula~\eqref{eq:TracyWidom1} to the sum over the permutations $\tau$ of the $n-1$ remaining integers. Since $\sg{\sigma} = (-1)^{j-1} \sg{\tau}$, we get
\begin{equation}
    T_{n,r} = \oint \left[
    \prod_{k=1}^n \frac{\dd z_k}{2 \I \pi z_k(1-z_k)} \ee^{t \left( z_k + \frac{1}{z_k} - 2 \right)}
    \right]
    \left[
    \prod_{k=1}^n \frac{z_1 \cdots z_k}{1- z_1 \cdots z_k}
    \right]
    \left[
    \prod_{k < l} \frac{z_k - z_l}{z_k z_l + 1 - 2z_k}
    \right]
    \sum_{j=1}^n
    \left[
    z_j^{r-1} (1-z_j) \prod_{i \neq j} \frac{z_i z_j + 1 -2z_j}{z_i - z_j}
    \right]
    \:.
\end{equation}
To further simplify this expression, we relabel the variables $z_k \to z_{n+1-k}$ and average the resulting expression over all the permutations $\sigma$ of the $n$ labels. This allows us to apply once again~\eqref{eq:TracyWidom1}, but to all the indices this time. We obtain
\begin{equation}
    T_{n,r} = \frac{1}{n!} \oint
    \left[
    \prod_{k=1}^n \frac{\dd z_k}{2 \I \pi (1-z_k)^2} \ee^{t \left( z_k + \frac{1}{z_k} - 2 \right)}
    \right]
    \left[
    \prod_{k \neq l} \frac{z_k - z_l}{z_k z_l + 1 - 2 z_k}
    \right]
    \left[
    \sum_{j=1}^n
    z_j^{r-1} (1-z_j) \prod_{i \neq j} \frac{z_i z_j + 1 -2z_j}{z_i - z_j}
    \right]
    \:.
\end{equation}
We now recognize another formula of Tracy and Widom~\cite{Tracy:2008b},
\begin{equation}
    \label{eq:TracyWidom2}
    \left[
    \prod_{k=1}^n \frac{1}{(1-z_k)^2}
    \right]
    \left[
    \prod_{k \neq l} \frac{z_k - z_l}{z_k z_l + 1 - 2 z_k}
    \right]
    = \det \left[
    \frac{1}{z_k z_l + 1 - 2 z_k}
    \right]_{1 \leq k,l \leq n}
    \:,
\end{equation}
which gives
\begin{equation}
    \label{eq:TnrInterm}
    T_{n,r} = \frac{1}{n!} \oint
    \left[
    \prod_{k=1}^n \frac{\dd z_k}{2 \I \pi}
    \right]
    \det \left[ \mathrm{K}_t(z_k,z_l) \right]_{1 \leq k,l \leq n}
    \left[
    \sum_{j=1}^n
    z_j^{r-1} (1-z_j) \prod_{i \neq j} \frac{z_i z_j + 1 -2z_j}{z_i - z_j}
    \right]
    \:,
\end{equation}
with $\mathrm{K}_t$ the kernel~\eqref{eq:KernelSEP}.

To simplify the sum over $j$, we introduce the matrix $C$ which has entries $C_{ij} = \frac{1}{z_i z_j + 1 - 2z_j}$ and the vector $a_j = \mathrm{K}_t(1,z_j) = \frac{1}{z_j -1}$. We define the vector $c$ solution of $a_i = \sum_j c_j C_{ji}$. Consider the rational fraction
\begin{equation}
    \sum_j \frac{c_j}{1- x z_j} - \frac{1}{1-x}
    = \frac{N(x)}{(1-x) \prod_j (1-x z_j)}
    \:,
    \quad
    N(x) = \sum_{j=1}^n (1-x) c_j \prod_{i \neq j}(1-x z_i) - \prod_j (1-x z_j)
    \:.
\end{equation}
By definition of $c$, $N(x = 2-z_i) = 0$ for all $i$. Since $N$ is a polynomial of degree $n$, is has to be of the form $N(x) = \alpha \prod_{i=1}^n[x - (2-z_i)]$. Evaluating $N$ at $x=1$, we deduce that $\alpha = (-1)^{n+1}$. Finally, evaluating $N$ at $x = 1/z_j$, we deduce that
\begin{equation}
    c_j = (z_j - 1) \prod_{k \neq j}
    \frac{z_k z_j - 2z_j + 1}{z_k - z_j}
    = \sum_i a_i [C^{-1}]_{ij}
    \:.
\end{equation}
From this relation, we can rewrite the sum over $j$ in~\eqref{eq:TnrInterm} in terms of $K_{ij} = \mathrm{K}_t(z_i,z_j) = \ee^{t \left(z_i + \frac{1}{z_i} - 2 \right)} C_{ij}$ as
\begin{equation}
    T_{n,r} = \frac{1}{n!}
    \oint
    \left[
    \prod_{k=1}^n \frac{\dd z_k}{2 \I \pi}
    \right]
    \det \left[ K_{kl} \right]_{1 \leq k,l \leq n}
    \left[
    -\sum_{i,j}
    z_j^{r-1} \ee^{t \left( z_j + \frac{1}{z_j} - 2 \right)} [K^{-1}]_{ij} a_i
    \right]
    \:.
\end{equation}
This can be written exactly as
\begin{equation}
    T_{n,r} = \omega \frac{\dd}{\dd s} \frac{1}{n!}
    \oint
    \left[
    \prod_{k=1}^n \frac{\dd z_k}{2 \I \pi}
    \right]
    \det \left[ K_{kl} + s \omega^{-1} a(z_l) b_{r;t}(z_k)  \right]_{1 \leq k,l \leq n}
    \Bigg|_{s=0}
    \:,
\end{equation}
with the functions $a$ and $b_{r;t}$ defined from~\eqref{eq:VectorsAB}. Summing over all values of $n$, we recognize the expansion of a Fredholm determinant~\cite{Tracy:2008b}, and thus,
\begin{equation}
    \avg{ \eta_r(t) \ee^{\lambda Q_t} }
    = \sum_{n \geq 1} \omega^{n-1} T_{n,t}
    = \frac{\dd}{\dd s}
    \det[ 1 + \omega \mathrm{K}_t + s b_{r;t} a^T] \Big|_{s=0}
    \:.
\end{equation}
Dividing by $\avg{\ee^{\lambda Q_t}} = \det[1 + \omega \mathrm{K}_t]$~\cite{Derrida:2009}, we obtain the expression~\eqref{eq:hrAsDet} given in the main text.

\textit{Appendix B: Derivation of the closed equation.---}
Our goal is to compute the double integral in Eq.~\eqref{eq:DblInteg0}. We can expand the term $\frac{1}{zz' + 1 -2z'}$ in powers of $z$, but that would generate positive powers of $z'$ and thus we cannot use the expression of $h_r$~\eqref{eq:Exprhr} which holds only for negative powers. We thus follow another approach and write $\frac{1}{zz' + 1 -2z'} = \frac{1}{z z'} \frac{1}{1 - (2z'-1)/(z z')}$. Expanding this term would generate negative powers of $z'$, but this can be done only for $|z| > \abs{\frac{2z'-1}{z'}}$. Therefore we deform the integration contour of $z$ from a small circle to a large one to perform this expansion. When doing so, the contour crosses the pole of the denominator, so we remove the residue
\begin{equation}
    \oint_{|z| \ll 1} \frac{\dd z}{2 \I \pi} \frac{\ee^{t \left( z + \frac{1}{z} - 2 \right)}}{z z'+1-2z'} z^{r-1}
    =
    - \frac{\left(2-\frac{1}{z'} \right)^{r-1}
    \ee^{t \left( \frac{z'}{2z'-1} - \frac{1}{z'} \right)}}{z'}
    + \oint_{|z| \gg \abs{\frac{2z'-1}{z'}}} \frac{\dd z}{2 \I \pi z z'} \frac{\ee^{t \left( z + \frac{1}{z} - 2 \right)}}{1 - \frac{2z'-1}{z z'}} z^{r-1}
    \:.
\end{equation}
Expanding the denominator we get
\begin{equation}
    \oint_{|z| \ll 1} \frac{\dd z}{2 \I \pi} \frac{\ee^{t \left( z + \frac{1}{z} - 2 \right)}}{z z'+1-2z'} z^{r-1}
    =
    - \frac{\left(2z'-1 \right)^{r-1}
    \ee^{t \left( \frac{z'}{2z'-1} - \frac{1}{z'} \right)}}{(z')^r}
    + \sum_{s \geq 1}
    p_{s-r} \sum_{m=1}^s (-1)^{m-1} \binom{s-1}{m-1} 2^{s-m} (z')^{-m}
    \:.
\end{equation}
Multiplying by $\ee^{t \left(z'+ \frac{1}{z'} - 2 \right)} \phi_{t;\omega}(z')$, we see that the first term on the r.h.s. does not have any pole at $z=0$ for $r \leq 0$, hence its integral on a small contour vanishes. Finally, after integration over $z'$ we get
\begin{equation}
\label{eq:IdentityProlong}
    \oint \frac{\dd z \dd z'}{(2 \I \pi)^2} \frac{\ee^{t \left(z + \frac{1}{z}-2 \right)}
    \ee^{t \left(z' + \frac{1}{z'}-2 \right)}}{z z' + 1 - 2z'}
    z^{r-1}
    \phi_{t;\omega}(z')
    = \sum_{s \geq 1}
    p_{s-r} \sum_{m=1}^s (-1)^{m-1} \binom{s-1}{m-1} 2^{s-m} h_{-m}
    \:,
\end{equation}
for any $r \leq 0$.

\textit{Appendix C: First correction to the large-scale behavior.---}
We derive Eq.~\eqref{eq:HydroCorrection} directly from the
microscopic closure. Set $\varepsilon=t^{-1/2}$,
$k=r$ for $r\geq1$ and $k=1-r$ for $r\leq0$, and
$x=\varepsilon(k-\frac12)$. Expand the common normalized
gradient as $\mathcal W_k=\varepsilon F(x)+\varepsilon^2H(x)
+O(\varepsilon^3)$. With $z=\varepsilon(s-\frac12)$,
\begin{align}
 p_k&=\varepsilon[G(x)+\tfrac\varepsilon2G'(x)]
 +O(\varepsilon^3),\nonumber\\
 p_{k+s}&=\varepsilon[G(x+z)+\varepsilon G'(x+z)]
 +O(\varepsilon^3),\nonumber\\
 (\Transf\mathcal W)_s
 &=\varepsilon\bigl[F(-z)+\varepsilon
 \{H(-z)+F'(-z)-zF''(-z)\}\bigr]+O(\varepsilon^3).
 \label{eq:hydroExpansions}
\end{align}
Here $G(x)=\ee^{-x^2/4}/\sqrt{4\pi}$, and $F(-z)$
is the continuation fixed by the microscopic reflection.
With $(\mathcal Kf)(x)=\int_0^\infty G(x+z)f(-z)\,\dd z$,
insertion into Eq.~\eqref{eq:ClosedMicro} and integration
by parts give
\begin{align}
 (1+\omega\mathcal K)F&=\omega G,\nonumber\\
 (1+\omega\mathcal K)H&=\tfrac\omega2G'
 +\omega\int_0^\infty zG''(x+z)F(-z)\,\dd z.
 \label{eq:hydroSubleading}
\end{align}
The differential operator
$\mathcal D=-[3x+(2+x^2)\partial_x+2x\partial_x^2]/4$
satisfies $\mathcal DG=G'/2$ and
\begin{equation}
 (\mathcal K\mathcal D-\mathcal D\mathcal K)F
 =\int_0^\infty zG''(x+z)F(-z)\,\dd z.
\end{equation}
Hence $H=\mathcal DF$: no further Wiener--Hopf equation
needs to be solved.

The physical density follows from the exact sum of the
normalized gradients,
$w_r=\rho_\pm+B_\pm\sum_{m\geq k}\mathcal W_m/
(1+\sum_{m\geq1}\mathcal W_m)$,
where $B_\pm=C_\pm-\rho_\pm$.
Set $I(x)=\int_x^\infty F(y)\,\dd y$ and
$J(x)=\int_x^\infty H(y)\,\dd y$.
The midpoint Euler--Maclaurin expansion yields
\begin{align}
 \sum_{m\geq k}\mathcal W_m
 &=I(x)+\varepsilon\bigl[J(x)+\tfrac12F(x)\bigr]
 +O(\varepsilon^2),\nonumber\\
 1+\sum_{m\geq1}\mathcal W_m
 &=c+\varepsilon J(0)+O(\varepsilon^2),
 \quad c=\sqrt{1+\omega}.
\end{align}

The known large-scale profile has the explicit
Laplace transform ~\cite{Grabsch:2022,Grabsch:2023}
\begin{equation}
\label{eq:HydroExplicit}
\int_0^\infty \ee^{-p x}
[\Phi_\pm(x)-\rho_\pm]\,\dd x
=\frac{B_\pm}{p}\left[
1-\frac{1}{c}\exp\left(
\frac{p}{\pi}\int_0^\infty
\frac{\log(1+\omega\ee^{-q^2})}
{p^2+q^2}\,\dd q
\right)\right],
\quad p>0.
\end{equation}
In particular, this solution gives the moments

\begin{equation}
 \int_0^\infty F(x)\,\dd x=c-1,\qquad
 \int_0^\infty xF(x)\,\dd x
 =-\frac{c}{\sqrt\pi}\operatorname{Li}_{1/2}(-\omega).
\end{equation}
Moreover, $H=\mathcal DF$ implies
$J(0)=-\frac14\int_0^\infty xF(x)\,\dd x$.
Using $\Phi_\pm(x)=\rho_\pm+B_\pm I(x)/c$
and $F(x)=-c\Phi_\pm'(x)/B_\pm$, we obtain
Eq.~\eqref{eq:HydroCorrection} without introducing
any additional unknown function.

\end{document}